\documentclass{article}

\usepackage[final, nonatbib]{neurips_2019}

\usepackage[utf8]{inputenc} % allow utf-8 input
\usepackage[T1]{fontenc}    % use 8-bit T1 fonts
\usepackage{hyperref}       % hyperlinks
\usepackage{url}            % simple URL typesetting
\usepackage{booktabs}       % professional-quality tables
\usepackage{amsfonts}       % blackboard math symbols
\usepackage[pdftex]{graphicx}
\usepackage{fancyhdr}
\title{Granular Motor State Monitoring of Free-Living Parkinson’s Disease Patients via Deep Learning}

\author{%
    Kamer A. Yuksel, Jann Goschenhofer, Hridya V. Varma, Urban Fietzek, Franz M.J. Pfister\\
    Dept. of Statistics \& Dept. of Neurology, Ludwig-Maximilians University; Munich, Germany\\
    \texttt{jann.goschenhofer@stat.uni-muenchen.de, urban.fietzek@med.uni-muenchen.de} \\
}

\begin{document}

\maketitle

\begin{abstract}
Parkinson's disease (PD) is the second most common neurodegenerative disease worldwide and affects around 1\% of the (60+ years old) elderly population in industrial nations. 

More than 80\% of PD patients suffer from motor symptoms, which could be well addressed if a personalized medication schedule and dosage could be administered to them. 
However, such personalized medication schedule requires a continuous, objective and precise measurement of motor symptoms experienced by the patients during their reglular daily activities. 
In this work, we propose the use of a wrist-worn smart-watch, which is equipped with 3D motion sensors, for estimating the motor fluctuation severity of PD patients in a free-living environment. 
We introduce a novel network architecture, a post-training scheme and a custom loss function that accounts for label noise to improve the results of our previous work in this domain and to establish a novel benchmark for nine-level PD motor state estimation.
\end{abstract}

\section{Introduction}

% intro and contrib
This paper is an extension of a previous work \cite{JannGoschenhofer2019} whithin which we compared the performances of classification, ordinal classification and regression on the motor fluctuation estimation problem using state-of-art time series classifcation (TSC) architectures. 
Therein, we concluded that the default FCN architecture in a regression setting performs best among several baseline methods.
Also, we found transfer learning from a simpler and weakly-labeled but also Parkinson's-related task an effective method to improve model performance further. 
In this current work, we improve both the model architecture and the customized performance measure.  
The primary contributions of this work over the current state of the literature are three-fold: the proposal of a new architecture for motor fluctuation estimation in a regression setting, the introduction of a post-training procedure to improve model generalization and the improvement of the custom loss to tackle the discreteness noise that is present in the labels given by clinical experts.

% literature overview - necessary?
Previous work in the area of PD often faces some practical limitations which we try to improve upon with our proposed method. 
Keijers et. al. \cite{KeijsersHorstinkGielen2003} were the first ones who proposed modelling approaches with a continuous response while the vast majority of the literature modelled this problem either as a binary (ON-OFF) or three-class classification with a discrete response about the motor fluctuation states of the patients. This was remarkable, as medical doctors need information about motor states on a more granular level in order to optimize the medication schedule and dosage. 
Moreover, only a minority of the literature used a LOSO validation scheme, which is crucial for real-life medical applications to ensure generalizability \cite{SaebLoniniJayaramanEtAl2017}; especially in this problem where there is a significant variety in how these symptoms affects patient motions. 
Finally, previous work often introduced impractical settings for the real-world usage in the PD treatment, which would not match the free-living conditions of patients such as the placement of several sensors over the body to instructing a limited set of specific actions to patients \cite{GhikaWiegnerFangEtAl1993, HoffVanDerMeerVanHilten2004, PastorinoCancelaArredondoEtAl2011, SalarianRussmannWiderEtAl2007}.
See \cite{JannGoschenhofer2019} for a more detailed literature overview.

\section{Architecture}

\begin{figure}[!h]
    \centering
    \includegraphics[width=1.0\linewidth]{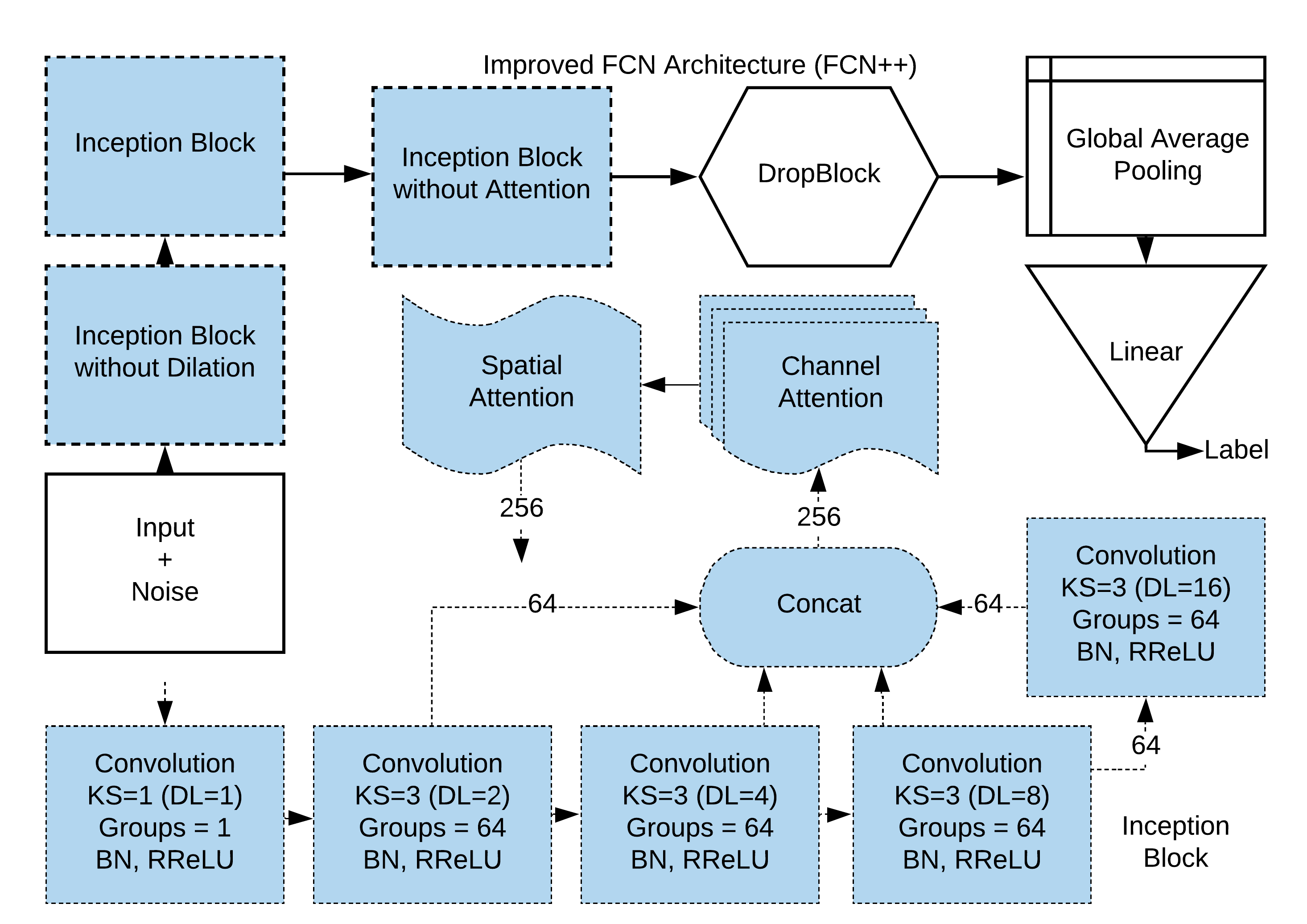}
    \caption{The improved FCN++ architecture (top) which includes Inception blocks (bottom) whith exponentially increasing dilations. Dilation is not used at the first and attention is not used at the last block. FCN+ is a samller version that is 2/3 of FCN++'s size as it does contain the channel attention.}
    \label{fig:1}
\end{figure}

In order to develop the improved version of the FCN architecture, which was the state-of-the-art in the TSC domain, this work was inspired from several versions of the Inception Networks \cite{SzegedyVanhouckeIoffeEtAl2016} that enable adaptive kernel-sizes for convolutional layers. 
Different from original Inception modules, we have allowed the architecture to also employ different dilation factors and hence the receptive fields adaptively. 
In the proposed Inception block (see Figure \ref{fig:1}), branches in every layer also have incremental levels of dilation in order to enlarge the receptive field similar to WaveNet \cite{Oord2016WaveNetAG} and TCN \cite{LeaVidalReiterEtAl2016} architectures, where it has previously been demonstrated that dilated convolutions support an exponential expansion of the receptive field without loss of coverage with less computational resources. 
The proposed Inception block allows the model to adapt its complexity in each layer via residual connections that converge to one of the four branches of different complexities \cite{DrozdzalVorontsovChartrandEtAl2016, HeZhangRenEtAl2015}. 
Branches that consist of multiple kernel-sizes and dilations allow the extraction of information from multiple resolutions and frequencies of the time series simultaneously. 
Attention modules come into play to enable the model to focus on the most important branches. These adaptive properties that have extended the FCN architecture also reduce its dependency on extensive hypertuning. 
Finally, we have utilized depth-wise separable convolutions to further reduce the number of computations and parameters. Grouped convolutions became popular due to their low computational cost and were shown to not just improve the speed of a model but also can improve its accuracy \cite{Xie2017AggregatedRT}.
In this work, we have employed Depthwise Convolutions, which are a special case of grouped convolutions where the number of groups is equal to the number of channels. 
Also, we decomposed ordinary convolutions into depthwise and pointwise convolutions as applied in Xception, MobileNet or ShuffleNet \cite{Chollet2017XceptionDL, Sandler2018MobileNetV2IR, Ma2018ShuffleNetVP}. 
In order to perceive an object, humans use a sequence of partial glimpses which can then be gathered to focus on the most important features of the object. 
Inspired by this idea, Woo et. al. \cite{Woo2018CBAMCB} introduced the channel and spatial attention modules, which have also been employed in this work. 
Given an intermediate feature map, their proposed attention module sequentially infers attention maps along two separate dimensions. 
Those maps are then multiplied with the input feature map for adaptive feature refinement. 
In this combined module, channel attention focuses on deciding 'what' the most meaningful features are among the channels (feature detectors) of the input feature map; the spatial attention focuses on finding 'where' the most informative regions of the object lies. 
Data augmentation and regularization techniques are crucial to improve the generalization capability of a method, especially in a LOSO cross-validation scheme. 
To achieve that, we have applied two data augmentation techniques, which have been shown to be useful in the PD domain \cite{UmPfisterPichlerEtAl2018}. 
The first method is the addition of noise to the input signal, which also aims to make the model robust to the input noise that is present especially in motion sensors. 
The second method divides the input signal into five blocks and then shuffles those blocks with a random permutation. 
In addition to data augmentation, we have employed DropBlock regularization at the last convolutional layer of the proposed architecture \cite{Ghiasi2018DropBlockAR} . 
As opposed to the widely-adopted Dropout method, spatially correlated units in a contiguous region of a feature map are dropped together in DropBlock, which enforces remaining units to learn enhanced features by preventing spatially correlated activations.

\section{Custom Loss}

One problem about collecting labels in a clinical environment is that labels given by clinicians are discretized and prone to subjectiveness. 
This situation is not always sufficiently consistent with the natural symptoms of patients, which are actually continuous, and thus often act as if a label-noise is added to the natural ground truth. 
Deep Learning methods are vulnerable to such noisy clinical labels as their overparametrization enables them to memorize these noisy samples during training. 
Since the motor severity of PD symptoms are labelled using discrete ratings, the labels corresponding to the similar natural motor states may vary significantly. 
The custom loss function is an adjusted version of the $L_2$ loss function, which takes this into account and adjusts for the discreteness noise within labels by assuming a fixed uncertainty $\beta$ in the quadratic error between the true label $y$ and the model prediction $\hat{y}$. 
Thus, it accepts all predictions that lie within this uncertainty bound. 
In addition, this custom loss was designed for the diagnosis of PD-related symptoms integrating the following clinical requirements \cite{JannGoschenhofer2019}: (1) Non-linearity of labels: misclassifying a sample with a label that is two levels away is more than twice as bad  as misclassifying by only one level (2) Asymmetry of labels: an exaggerated diagnosis in the true pathological direction should be better than an opposing diagnosis in the wrong pathological direction, even if the label difference between them is the same and (3) Misclassification cost: the cost of misclassifying a patient in the wrong direction is proportional to the motor severity.

\begin{equation}
L^{custom} (y, \hat y, \alpha = 0.25, \beta = 0.25) = \left( \alpha + sign (\hat y - y) \right)^2 \cdot max \left(0, \left( \hat y - y \right)^2 - \beta \right)
\end{equation}

where $y$ is the true label, $\hat{y}$ is the model prediction, and $\alpha$ and $\beta$ are parameters that control costs defined for the asymmetry and the allowance of the label relaxation (r) in the custom loss, respectively. 
Finally, a weighted average of losses for each sample in the mini-batch are taken according to inverse-frequency of each class [-4in the dataset to prevent a bias due to its high-class imbalance. 
This class-weighting scheme is also adopted for the evaluation metrics in the experimental results.

\section{Experiments}

% data and label description
The motor fluctuation severity dataset contains on average six hours recording of the 3D accelerometer and gyroscope values collected from 39 PD patients using a single Microsoft Band 2 fitness-tracking smart-watch, which was mounted on the wrist of their dominant hand. 
The real-world applicability was taken into great consideration while selecting the quantity and placement of the motion sensor as opposed to several previous research in this domain. 
In order to augment the training set, a sliding window is utilized while dividing its time series into one-minute bins with 80\% overlap between neighbouring windows. 
After down-sampling from 62.5 Hz to 20 Hz, which is sufficient to analyze PD symptoms, Euclidean norms of the 3D raw signals from both sensors are extracted as input features for each one-minute bin, which was assigned a motor fluctuation severity label as explained in the following paragraph. 
The Euclidean norms are rotation invariant features, which are not affected by the way the device is worn. 
Extracted features are normalized using the uniformly-distributed version of the Quantile Transformer in Scikit-Learn. 

The severity of PD symptoms was assessed by a clinical rater using a combination of the MDS-UPDRS \cite{GoetzTilleyShaftmanEtAl2008} and the mAIMS \cite{LaneGlazerHansenEtAl1985}, which are ratings based on a diagnostic questionnaire where medical doctors rate the severity of bradykinesia and dyskinesia respectively, on a discrete scale of 0 - 4, with 0 being the ON state. 
When both ratings are combined, each minute of motion data is labelled with a rating between -4 and 4 (from bradykinetic to dsykinetic state with 0 being the ON state) by a doctor who monitored PD patients in a free-living setting. 
Labels assigned for the minute-wise motor fluctuation severities of patients were highly imbalanced as consistent with the nature of the PD. 
This high class imbalance was taken into account on both the methodologies and evaluation metrics utilized as suggested by \cite{GengLuo2018, HeGarcia2009}. In addition to that, we make use of a second unsupervised dataset which did not have labels for motor symptom fluctuation severities but only the meta-information on whether the recording belongs to one of 97 PD patients or of 39 healthy controls. 
This second dataset, is used for pre-training of the supervised model \cite{YosinskiCluneBengioEtAl2014, LaengkvistKarlssonLoutfi2014} via an auxiliary binary classification task (PD or healthy) in order to improve the performance on the supervised regression task. 
This way, we could leverage the unsupervised data from a relatively much higher amount of 136 subjects. 

% training specs and validation
For validating the generalization capability of the proposed method, we have employed Leave-one-subject-out (LOSO) validation and selected the evaluation epoch for each setting based on the minimum average loss of the individual folds. 
Experiments were ran on a Nvidia Tesla V100 GPU using an improved version of the Adam optimizer for a better generalization performance via a decoupled weight decay \cite{LoshchilovHutter2017} of 5e-6 and a partially adaptive momentum \cite{Chen2018ClosingTG} of 0.4. 
In all experiments, a batch-size of 128, a learning rate of 1e-4 and an epsilon parameter of 5e-3 was set. 
The standard deviation of the additive Gaussian input noise was set to $0.05$ and a DropBlock rate of 20\% is used in the last convolutional layer on the experiments if applied. 
Predictions are post-processed using a Gaussian smoothing over the neighbouring windows before calculating the regression metrics in Tables 1-4. 
The regression output is rounded for mapping them back to the class labels for the calculation of classification metrics. 
Due to the high-class imbalance in the dataset, the class-weighted versions of all evaluation metrics are used as results would not be interpretable otherwise\cite{BrodersenOngStephanEtAl2010}. 

% posttraining
Finally, a post-training strategy was used which was introduced by Han et. al. \cite{Han2017DSDDT}. 
This strategy starts with a fully trained dense model and then regularizes it by model sparsification to achieve better optimization performance. 
This is done by sorting the weights of the network, pruning low-weight connections and post-training the resulting sparse network. 
In their method, the pruned weights are re-introduced and initialized to zero before training the complete dense network again with a smaller learning rate. 
In this work, we have only applied the first two phases by continuing on sparse training on the 75\% pruned connections in each convolutional layer after the initial convergence with the dense network. 
This post-training allowed for the significant improvement of results achieved from the first convergence with the dense network due to the large kernel sizes of FCN and redundancy of Inception branches in other models.

% results
Tables 1-4 show the results of several experiments conducted for FCN, FCN+ and FCN++ models using different training schemes. 
As the heavily under-represented -4 and 4 boundary classes performed significantly worse than others, 7-class classification results are also reported for experiments. 
The proposed custom loss consistently improved (for both 7 and 9 class) classification metrics \cite{SokolovaLapalme2009} almost for all of the models and schemes when its relaxation (r) is enabled. 
Whereas, the non-relaxed version of the custom loss performed better with the regression metrics including the relaxed accuracy, which has similar characteristics with the regression metrics due to its prediction tolerance of $\pm$1 class. 
The proposed pre-training and post-training schemes were useful for almost all models both when they were applied individually and as a combination. 
Except for the last training scheme, where both pre and post training was applied, the proposed FCN+ and FCN++ architectures consistently outperformed the default FCN architecture. 
The FCN++ architecture was better than the FCN+ architecture in all of the cases where pre-training was applied due to its larger number of parameters (54.3k vs 37.4k). 
The default FCN architecture outperformed all other architectures in all regression metrics and the relaxed accuracy when both pre- and post-training is applied. 
This may be due to its four-times larger number of parameters (265k) that contain the largest amount of redundant weights and thus benefits the most from post-training. 
The direct comparison of the results with most of the related literature work was difficult due to their lack of LOSO cross-validation as well as class-weighted evaluation metrics, which are both crucial for evaluation of their methodologies. 
The best accuracies reported in the literature with a single sensor on a free-living environment using a LOSO validation scheme are 73.9\% \cite{HssayeniBurackGhoraani2016} for two-class and 63.1\% \cite{UmPfisterPichlerEtAl2018} for three-class motor state classification problems. 
The method proposed in this work yields a relaxed accuracy of 80.0\% on a nine-class setting. 
As further demonstrated in Table 4, the proposed method clearly succeeded in establishing a new benchmark for this problem by achieving comparable classification metrics on a 7/9-class problem to what the previous work obtained on a 2/3-class setting, despite the primary objective of the model training was actually achieving better regression metrics (especially in terms of the proposed custom loss).

\section{Conclusion}

In this work, we have proposed the use of a wrist-worn smart-watch, which is equipped with 3D motion sensors, for estimating the motor fluctuation severities of PD patients in their free-living conditions. 
Deep Learning techniques enabled us to outperform the previous work which often also had practical limitations for the real-world applications. 
The primary contributions of this work over the previous work in the literature were as follows: (1) proposing a state-of-the-art architecture for the motor fluctuation estimation using a regression setting (2) proposing a pre-training technique for improving the generalization of the model to previously unseen patients while training with a limited clinically labelled dataset (3) proposing a relaxed loss function with the aim of tackling the discreteness noise by the labels given by clinical raters. 
Experimental results shown that accurate estimation of nine-level motor fluctuation severity of PD patients, which is one of the most relevant constructs for the clinical decision taking in PD, is possible on the comfort of their home. 
This accurate estimation can enable optimizing the medication schedule and dose in such a way that the patient has an improved chance to spend the entirety of his waking day in normal state; in other words, achieving a stable ON condition while impeding OFF or dyskinetic motor states. 
The personalized medication, which should be adapted as the disease progresses, can allow improving the life of PD patients who are suffering from severe symptoms just because of the imperfect medication.

\newpage

\section*{Appendix}

\begin{table}[!h]
    \centering
    \caption{Experimental results without pre- and post-training (pp)}
    \begin{tabular}{llllllll}
    	\toprule
    	\textbf{Model} & \textbf{MAE} & \textbf{MSE} & \textbf{Acc-7} & \textbf{Acc-9} & \textbf{F1-7} & \textbf{F1-9} & \textbf{Acc$\pm$1} \\ 
    	\toprule
    	FCN & 1.010 & 1.713 & 0.365 & 0.307 & 0.362 & 0.356 & 0.701 \\
    	\midrule
    	FCN (r) & 0.983 & 1.478 & 0.384 & 0.292 & 0.404 & 0.392 & 0.703 \\
    	\midrule
    	FCN+ & 0.940 & 1.449 & \textbf{0.390} & \textbf{0.329} & \textbf{0.419} & \textbf{0.404} & \textbf{0.747} \\
    	\midrule
    	FCN+ (r) & \textbf{0.908} & \textbf{1.362} & 0.388 & 0.310 & 0.411 & 0.397 & 0.746 \\
    	\midrule
    	FCN++ & 0.960 & 1.458 & 0.370 & 0.297 & 0.397 & 0.388 & 0.711 \\
    	\midrule
    	FCN++ (r) & 0.973 & 1.550 & 0.378 & 0.300 & 0.387 & 0.374 & 0.722 \\
    	\bottomrule                          
    \end{tabular}	
    \vspace{0.5cm}
    \caption{Experimental results with pre-training}
    \begin{tabular}{llllllll}
    	\toprule
    	\textbf{Model} & \textbf{MAE} & \textbf{MSE} & \textbf{Acc-7} & \textbf{Acc-9} & \textbf{F1-7} & \textbf{F1-9} & \textbf{Acc$\pm$1} \\ 
    	\toprule
    	FCN & 1.031 & 1.658 & 0.373 & 0.272 & 0.358 & 0.338 & 0.693 \\
    	\midrule
    	FCN (r) & 1.055 & 1.692 & 0.330 & 0.264 & 0.366 & 0.358 & 0.657 \\
    	\midrule
    	FCN+ & 0.951 & 1.409 & 0.375 & 0.301 & 0.394 & 0.383 & 0.718 \\
    	\midrule
    	FCN+ (r) & 1.000 & 1.617 & 0.368 & 0.282 & 0.387 & 0.373 & 0.727 \\
    	\midrule
    	FCN++ & \textbf{0.913} & \textbf{1.304} & 0.380 & 0.287 & 0.388 & 0.372 & \textbf{0.760} \\
    	\midrule
    	FCN++ (r) & 0.945 & 1.426 & \textbf{0.382} & \textbf{0.302} & \textbf{0.404} & \textbf{0.393} & 0.732 \\
    	\bottomrule                          
    \end{tabular}	
    \vspace{0.5cm}
    \caption{Experimental results with post-training}
    \begin{tabular}{llllllll}
    	\toprule
    	\textbf{Model} & \textbf{MAE} & \textbf{MSE} & \textbf{Acc-7} & \textbf{Acc-9} & \textbf{F1-7} & \textbf{F1-9} & \textbf{Acc$\pm$1} \\ 
    	\toprule
    	FCN & 0.999 & 1.454 & 0.342 & 0.256 & 0.408 & 0.397 & 0.704 \\
    	\midrule
    	FCN (r) & 1.047 & 1.636 & 0.362 & 0.276 & 0.391 & 0.378 & 0.687 \\
    	\midrule
    	FCN+ & 0.981 & 1.474 & 0.\textbf{404} & \textbf{0.307} & \textbf{0.440} & \textbf{0.424} & 0.695 \\
    	\midrule
    	FCN+ (r) & 1.033 & 1.561 & 0.334 & 0.255 & 0.379 & 0.372 & 0.666 \\
    	\midrule
    	FCN++ & \textbf{0.907} & \textbf{1.321} & 0.365 & 0.305 & 0.396 & 0.385 & \textbf{0.758} \\
    	\midrule
    	FCN++ (r) & 0.949 & 1.439 & 0.367 & 0.299 & 0.404 & 0.396 & 0.723 \\
    	\bottomrule                          
    \end{tabular}	
    \vspace{0.5cm}
    \caption{Experimental results with pre- and post-training (pp)}
    \begin{tabular}{llllllll}
    	\toprule
    	\textbf{Model} & \textbf{MAE} & \textbf{MSE} & \textbf{Acc-7} & \textbf{Acc-9} & \textbf{F1-7} & \textbf{F1-9} & \textbf{Acc$\pm$1} \\ 
    	\toprule
    	FCN & \textbf{0.894} & \textbf{1.227} & 0.378 & 0.282 & 0.347 & 0.327 & \textbf{0.800} \\
    	\midrule
    	FCN (r) & 0.897 & 1.293 & 0.394 & 0.313 & 0.411 & 0.393 & 0.751 \\
    	\midrule
    	FCN+ & 1.020 & 1.630 & 0.352 & 0.282 & 0.368 & 0.363 & 0.677 \\
    	\midrule
    	FCN+ (r) & 1.001 & 1.554 & 0.373 & 0.292 & 0.384 & 0.370 & 0.721 \\
    	\midrule
    	FCN++ & 0.934 & 1.298 & 0.357 & 0.277 & 0.384 & 0.372 & 0.731 \\
    	\midrule
    	FCN++ (r) & 0.911 & 1.350 & \textbf{0.405} & \textbf{0.317} & \textbf{0.432} & \textbf{0.419} & 0.742 \\
    	\bottomrule                          
    \end{tabular}	
    \label{table:all}
\end{table}

\end{document}